\documentclass{aa}
\usepackage{graphicx}
\usepackage{xcolor}
\usepackage{txfonts}
\usepackage[draft]{hyperref}
\usepackage{placeins}

\newcommand{\vvcrit}{$v/v_\mathrm{crit}\ $}
\newcommand{\nsample}{490}

\begin{document}

\title{Age uncertainties of red giants due to cumulative
  rotational \\mixing of progenitors calibrated by asteroseismology
  \thanks{The full Table A.1 is only available in electronic form at the CDS via anonymous ftp to cdsarc.u-strasbg.fr (130.79.128.5) or via http://cdsweb.u-strasbg.fr/cgi-bin/qcat?J/A+A/}
  }

   \subtitle{}

   \author{D. J. Fritzewski\inst{1}
          \and C. Aerts\inst{1,2,3}
          \and J. S. G. Mombarg\inst{4}
          \and S. Gossage\inst{5}
          \and T. Van Reeth\inst{1}
          }

   \institute{Institute of Astronomy, KU Leuven, Celestijnenlaan 200D, 3001, Leuven, Belgium\\
   	\email{dario.fritzewski@kuleuven.be}
      \and
      Department of Astrophysics, IMAPP, Radboud University Nijmegen,
      PO Box 9010,
      6500 GL Nijmegen, The Netherlands
      \and
      Max Planck Institut für Astronomie, Königstuhl 17, 69117 Heidelberg, Germany
      \and
      IRAP, Universit\'e de Toulouse, CNRS, UPS, CNES, 14 avenue
      \'Edouard Belin, F-31400 Toulouse,
      France
      \and
      Center for Interdisciplinary Exploration and Research in
      Astrophysics (CIERA),
      Northwestern University, 2145 Sheridan Road, Evanston, IL 60208, USA
   }

   \date{}


  \abstract
   {Galactic archaeology largely relies on precise ages of distant evolved stars in
   the Milky Way. Nowadays, asteroseismology can deliver ages for many
   red giants observed with high-cadence, high-precision
   photometric space missions such as CoRoT, {\it Kepler}, K2, TESS, and
   soon PLATO.}
   {Our aim is to quantify age uncertainties of currently slowly-rotating red giants due to the cumulative effect of their fast rotation
     during core-hydrogen burning. Their rotation
     in earlier evolutionary phases
     caused mixing of elements resulting in heavier helium cores
     and the prolongation of their main sequence lifetime. These rotational
     effects are usually ignored when age-dating red giants, despite our knowledge of fast rotation for stars with  $M\ge 1.3$\,M$_\odot$.}
   {We use a sample of \nsample{} F-type gravito-inertial pulsators ($\gamma\,$Doradus stars) with precise asteroseismic estimates of
     their internal rotation rate from {\it Kepler\/} asteroseismology
     and with luminosity estimates from \emph{Gaia}. For this sample, which includes stars
     rotating from nearly zero to
about 60\,\% of the critical rate, we compute the
     cumulative effect on the age in their post-main sequence evolution caused by rotational mixing on the main sequence. We use stellar
     model grids with different physical prescriptions mimicking
     rotational mixing to assess systematic uncertainties on the age. }
   {With respect to non-rotating models, the sample of \nsample{} $\gamma\,$Doradus stars, as red giant progenitors, reveals
     age differences up to 5\,\% by the time they start
     hydrogen-shell burning when relying on the theory of
     rotationally induced diffusive mixing as included in the MIST isochrones.
     Using rotational mixing based on an advective-diffusive approach including meridional circulation leads to an age shift of 20\,\% by the time of  the tip of the red giant branch.}
   {Age-dating of red giants is affected by the cumulative effect of
     rotational mixing during the main sequence.
Such rotationally-induced age shifts should be taken into
     account in addition to other effects if the aim is to
     perform Galactic archaeological studies at the highest precision.}


   \titlerunning{Age-dating red giants from asteroseismically calibrated
     rotational mixing of their progenitors}
\authorrunning{Fritzewski et al.}

   \maketitle
%

   \section{Introduction}

Transport processes in stellar interiors have a significant
impact on the global properties and the chemical evolution of stars
across the entire mass range \citep[e.g.,][for high-, \mbox{intermediate-,}
  and low-mass stars,
  respectively]{Martins2013,Pedersen2021,Lagarde2017}. Asteroseismology
has pinpointed two dominant tightly connected effects of chemical mixing due to the accumulation over the core hydrogen burning main-sequence phase: a
considerable increase in the main-sequence duration and a heavier
helium core at the end of the main sequence \citep[up to a factor two
  more massive than without envelope mixing,
  see][]{Johnston2021,Pedersen2022a}. In this work, we
consider the former for intermediate-mass stars from the
viewpoint of age-dating red giant stars for Galactic archaeology.

Unravelling the formation and evolutionary history of the Milky Way
has many facets. One important element concerns chemical tagging of
stars dispersed throughout the galaxy
\citep[e.g.,][]{DeSilva2015,Schiavon2017,Vitali2024}. Other aspects involve
determination of the spatial structure and evolution of stellar
populations, moving groups, and streams in the Milky Way's bulge, bar,
thin, and thick discs
\citep[e.g.,][]{Bovy2016,Rojas2017,SilvaAguirre2018,Helmi2018}.  Since about a
decade, asteroseismology of red giants offers high-precision ages for these
old bright stars across the Milky Way \citep[see the extensive
  studies by][]{Miglio2013,Chiappini2015,Montalban2021,Hon2021,Wang2023}.
Age-chemo-kinematic properties of red giants indicate efficient radial
migration in the thin disc, different chemo-dynamical histories of the
thick and thin discs, and changes in the star formation after the
formation of the thick disc \citep{Miglio2021}.  The age
determinations of these populations of red giants in the Galaxy are of
critical importance to ensure a proper interpretation of the archaeological sites of the
Milky Way. While various core boundary mixing prescriptions are nowadays
being considered for the core helium burning phase of red giants \citep[e.g.,][]{Noll2024}, almost none of the constructed models take into account the previous long history of these stars during the core hydrogen phase, determining the size and mass of the helium core. Yet, the  effective chemical mixing in the progenitor stars needs to be treated with great care \citep{Montalban2013}.

In this work, we focus on well-studied intermediate-mass gravity-mode pulsators on the
main sequence. These $\gamma\,$Doradus ($\gamma\,$Dor)
stars are the progenitors of red giants with a mass above about
1.3\,M$_\odot$. Starting from the measured internal rotation rates of a
population of $\gamma\,$Dor stars covering the mass range from
1.3\,M$_\odot$ to 2\,M$_\odot$, we aim to study the arising systematic age uncertainties of red giants, when ignoring the rotation of their progenitors in their core hydrogen burning phase.

For red giants of lower mass, it is
justified to ignore the mixing caused by rotation, because their
dwarf progenitors experience efficient rotational slow-down due to
magnetic braking \citep{Barnes2007}, as confirmed by gyrochonology
studies of old open clusters based on high-precision space photometry
\citep{Meibom2011a,Meibom2011b,Meibom2015,Barnes2016}.
Even if magnetic braking weakens during the second half of the main
sequence \citep{vanSaders2016,Hall2021}, stars born with a mass below
1.2\,M$_\odot$ are slow rotators from early on in their evolution,
their rotation period being far longer than their critical break-up
period. In addition, the rotational evolution of cool stars is only a function of mass (and metallicity), thus suppressing possible age spreads due to rotation. Hence, we only consider red giants with a mass above
1.2\,M$_\odot$ in this work and study the impact of their
progenitors' rotation on their age-dating for the evolutionary phases
beyond the main sequence.

\section{Sample selection and global stellar parameters}
\label{sec:model}

A large homogeneous sample of 611 $\gamma$\,Dor stars with
high-precision measurement of their near-core rotation frequency,
$\Omega_{\rm rot}$, from {\it Kepler\/} space photometry \citep{Borucki2010} was presented
in \citet{GangLi2020}.  The other asteroseismic observable deduced
from their detected series of identified low-degree modes of
consecutive radial order is the buoyancy travel time,
$\Pi_0$, defined as
\begin{equation}
\Pi_0 \equiv 2\pi^2 \left(\int_{r_1}^{r_2} \frac{N(r)}{r} {\rm d}r\right)^{-1} ,
\end{equation}
where $N(r)$ is the Brunt-Väisälä frequency and $r_1$ and $r_2$ are the inner and outer position of the mode
propagation cavity \citep{Aerts2010}. This observable is a
measure of the size and shape of the gravity-mode cavity, which changes
during the evolution of the star. The
chemical gradient profile notably changes as the stars' convective cores shrink throughout
the main sequence. The resulting change in the shape of $N(r)$ has been measured from
gravito-inertial asteroseismology based on 4-year {\it Kepler\/} light
curves \citep[e.g.,][]{Aerts2021-GIW}.

We use these two powerful asteroseismic observables, $\Omega_{\rm rot}$
and $\Pi_0$, characterising the deep stellar interior of $\gamma\,$Dor
stars to estimate the
stellar masses and radii from grid modelling. Such type of asteroseismic
modelling requires additional constraints to break degeneracies and to find unique solutions. Since we wish to apply ensemble
asteroseismic grid modelling, we need to use a homogeneous set of
global stellar parameters. In contrast to \citet{GangLi2020} who used {\it Kepler\/} input catalogue values, we
extract stellar parameters from \emph{Gaia\/} Data Release 3
\citep[DR3,][]{Vallenari2023,Creevey2023b}, which are appropriate for F-type
$\gamma\,$Dor pulsators as shown by \citet{Aerts2023}. This yields
astrophysical parameters based on the \emph{Gaia} General Stellar Parametrizer from Photometry (GSP-Phot) for 536 of the 611 stars. Six of these
objects are removed as they are known binaries with pulsating
components so their parameters in the Gaia data might be inaccurate.

\begin{figure}
    \includegraphics[width=\columnwidth]{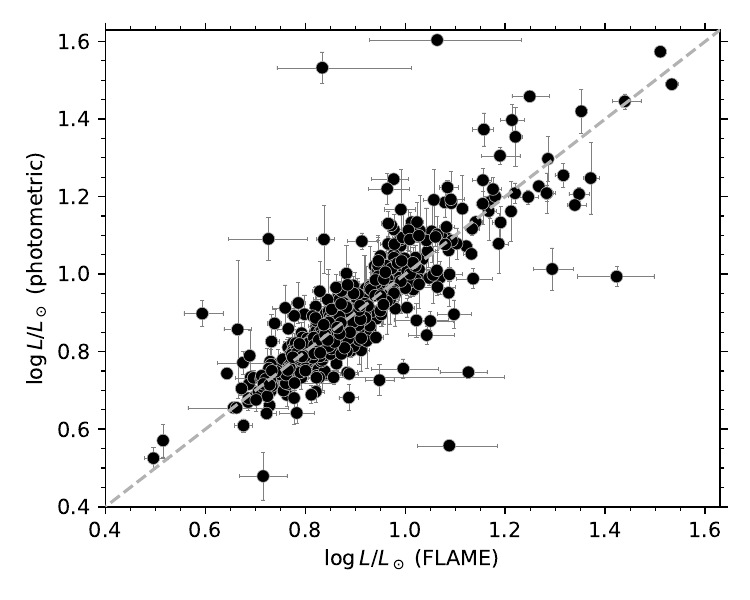}
    \caption{Comparison between luminosity values obtained from the \emph{Gaia} absolute $G$ magnitude (photometric luminosity) and the model dependent \emph{Gaia}-FLAME. The dashed line is the bisector.}
    \label{fig:spec_values}
\end{figure}

\citet{Mombarg2021} have shown the benefit of using the \emph{Gaia} stellar
luminosity ($L/L_\sun$), along with $\Pi_0$, the surface gravity ($\log g$)
and the effective temperature ($T_{\rm eff}$), in
asteroseismic grid modelling of identified prograde dipole modes for a sample of 37 $\gamma\,$Dor stars
with high-resolution high signal-to-noise spectroscopy. Due to the very uncertain $\log g$ estimates, we perform grid modelling from
the method by \citet{Mombarg2019} but based only on the three observables $\Pi_0$,
$T_{\rm eff}$, and $L/L_\odot$.

At least two ways occur to obtain the
stellar luminosity from \emph{Gaia\/} DR3.  First the {\it Gaia\/} astrophysical
parameters table includes the luminosity estimate from the Final Luminosity Age Mass Estimator \citep[FLAME,][]{Creevey2023b}.  Secondly, we calculate it from the
observed $G$-band magnitude using the parallax, extinction estimate, and
bolometric correction \citep[see][for the \emph{Gaia} bolometric correction
  computation]{Creevey2023b}.  The former method relies on the same
input parameters as the latter, but includes a rather sophisticated
scheme of placing these values on theoretical isochrones to deduce
$\log\,L/L_\odot$.  These isochrones are calibrated from slowly rotating
benchmark stars, while $\gamma\,$Dor stars tend to be faster rotators. Hence
the FLAME luminosities, while seemingly precise, may suffer from
systematic uncertainties.  In order to assess this, we compute the
luminosities without relying on stellar models.
The downside of this method is that we need an estimate of the
interstellar reddening (which we take from {\it Gaia\/} DR3). Yet, for most stars this reddening is small ($E(G_\mathrm{BP})-E(G_\mathrm{RP})<0.25$\,mag) as they are positioned away from the Galactic plane and typically nearby at a distance $d<1.5$\,kpc.
Comparing the two estimates allows us to evaluate the involved uncertainties.

Figure\,\ref{fig:spec_values} provides a comparison between both values and reveals that they are in good agreement with each other. The scatter of $\pm 0.1$\,dex around the line of unity is expected because the photometric luminosities include fewer and independently treated variables whereas FLAME treats all variables in a global framework with additional model constraints. As seen in Fig.\,\ref{fig:spec_values}, we find some obvious outliers to the
generally tight correlation, with the photometric luminosities
maximally up to a factor of five higher than the FLAME values.

To further clean our sample, we investigate the correlation between
the effective temperatures given by \citet{GangLi2020} from the {\it
  Kepler\/} Input Catalogue and those in {\it Gaia\/} DR3. These temperatures
are mostly in agreement. However, 19 stars have an effective
temperature outside the $\gamma\,$Dor range, rather pointing to
them being early A-type or B-type pulsators ($T_{\rm eff}>8500$\,K).
We use the same approach as \citet{Aerts2023} to reclassify these
gravity-mode pulsators as Slowly Pulsating B stars.

Even in this cleaned sample some stars exhibit larger
$\Pi_0$ values than those expected for $\gamma$\,Dor
stars \citep{VanReeth2016}. Since we will employ grid-based
asteroseismic modelling, we limit
the sample to stars with $\Pi_0<6500$\,s. This value might exclude some of the youngest stars but it ensures that the
modelled stars fall onto the used grid.
Our final sample includes \nsample{} stars with homogeneously deduced  astrophysical parameters from {\it Gaia\/} DR3 and 4-year {\it
  Kepler\/} light curves.

\begin{figure}
    \includegraphics[width=\columnwidth]{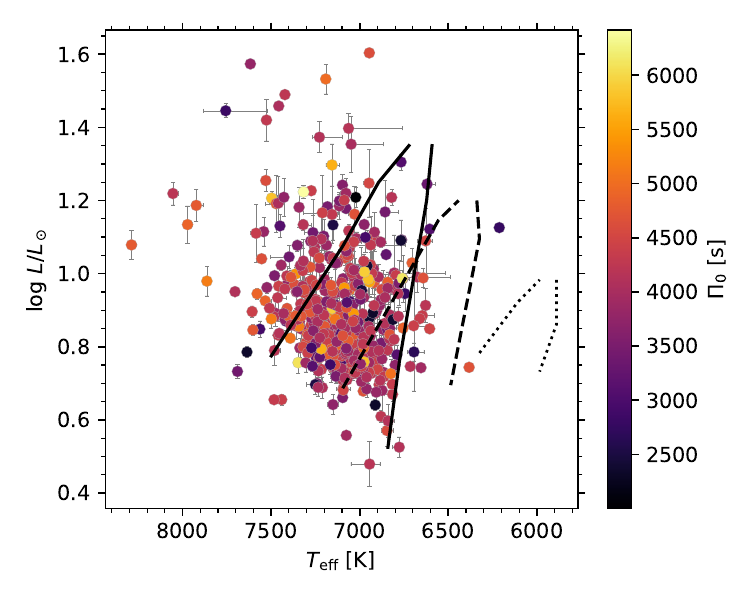}
    \caption{Hertzsprung-Russell diagram of our final sample containing \nsample{} $\gamma$\,Dor stars. The colour-coding gives the $\Pi_0$ which is used as an input parameter in the grid modelling. The solid lines indicate the theoretical $\gamma$\,Dor instability strip from \cite{Dupret2005} with a mixing-length parameter $\alpha_\mathrm{MLT}=2$, while the dashed and dotted lines show the position of the instability strip for $\alpha_\mathrm{MLT}=1.5$ and $\alpha_\mathrm{MLT}=1$, respectively.}
    \label{fig:HRD}
\end{figure}

We show the final sample in the observational Hertzsprung-Russell diagram (HRD) in Fig.~\ref{fig:HRD} along with the measured $\Pi_0$. Further, we show the theoretical $\gamma$\,Dor instability strips from \cite{Dupret2005}. The lower edge of the observed instability strip agrees very well with theoretical predictions (with a mixing-length parameter $\alpha_\mathrm{MLT}=2$), whereas the upper edge is at higher luminosities \citep[cf.\, also ][]{Mombarg2024}. These three instability regions were calculated for only one choice of input physics and for one excitation mechanism. Many {\it Gaia\/} DR3 candidate $\gamma$\,Dor stars are observed at higher luminosities \cite[e.g.][]{DeRidder2023} as we also find here.

\section{Asteroseismic parameter estimation}
\label{sec:astero_para}

Our nominal asteroseismic modelling purpose for this work is to
estimate each star's mass ($M$) and evolutionary stage represented by
the core hydrogen mass fraction ($X_c$).  In order to achieve this,
our approach consists of an MCMC-based grid search for each of the
\nsample{} stars in our final sample which we model
independently as it concerns field stars not belonging to an
astrophysical ensemble. We used the measured $\Pi_0$ along with {\it Gaia\/}
DR3 $L/L_\sun$ and $T_{\rm eff}$ estimates as inputs to place the stars on a
stellar evolution grid.

The grid consists of quasi-randomly sampled 1-dimensional (1D) Modules for Experiment in Stellar Astrophysics \cite[MESA r11701,][]{Paxton2011, Paxton2013, Paxton2015, Paxton2018, Paxton2019} models computed by
\citet{Mombarg2021}, to which we refer for the details of the input
physics. In terms of element transport, we recall here that the
overall chemical mixing caused by various active physical phenomena
(such as rotational mixing, internal gravity wave mixing, magnetic
diffusion, etc.) is
simplified in this grid of 1D models in terms of 1) instantaneous full convective
mixing in the core relying on the mixing-length theory
\citep{BohmVitense1958}, 2) diffusive exponentially decaying
overshooting with one free parameter ($f_{\rm ov}\in [0.01,0.03]$) in
the transition layer between the convective core and the radiative
envelope, and 3) constant envelope mixing described by a free
parameter. This approach allows us to assess the level of mixing in
these three zones without having to rely on uncalibrated theories of
element transport, because these 1D approximations of inherently 3D
phenomena result in
unsuitably spiky mixing profiles throughout the stellar interior
making pulsation computations unreliable \citep[cf.\,][for more
  detailed motivations of this commonly adopted approach in
  asteroseismic modelling of stars with a convective
  core]{Aerts2021-RMP}.  The covered mass range of the grid is
$[1.3,2.0]\,$M$_\odot$, while the metallicity $Z$ varies between 0.011
and 0.023 following the measured values for the sub-sample of 37 best
characterised $\gamma\,$Dor stars with high-precision spectroscopy and
its homogeneous analysis \citep{VanReeth2015}.

For our MCMC sampling, we use flat priors in mass ($M$), core-hydrogen content ($X_c$), and core overshoot ($f_\mathrm{ov}$) while the prior in metallicity ($Z$) is
taken from a Gaussian distribution centred on $Z_0=0.014$
($\sigma=0.01$), given that the stars are young from a Galactic perspective and hence are expected to have near Solar metallicity. This is indeed found from various studies based on high-precision spectroscopic analyses done for various subsets of the brighest $\gamma\,$Dor stars in the sample, as summarised by \citet{Gebruers2021}.
Moreover, given its inferior
effect on forward asteroseismic modelling based on the fitting of
actual identified mode frequencies \citep{Aerts2018} compared to the four unknowns
$(M,Z,f_{\rm ov},X_c)$ found by \citet{Mombarg2021}, and to reduce
dimensionality, we fix the envelope mixing to the lowest value of
$D_\mathrm{mix}=1\,$cm$^2$\,s$^{-1}$ in the grid.  This effectively
comes down to the assumption that rotational mixing is mainly active in
the shear layer above the convective core but has negligible effect in
the outer envelope of rotating F-type stars.  We come back to this
assumption in Sect.\,\ref{sec:systematics}, where we assess the effect of higher levels of
rotational mixing outside the core boundary layer
on asteroseismic parameter determinations. The merit function for the grid evaluation is the Euclidean distance between the grid points and the measurements.

For each star the grid search results in a distribution of $M$,
$X_c$, $f_\mathrm{ov}$, and $Z$. The latter is
restricted by the chosen prior but still varies across the whole range
of the model grid for some stars, showing the need for these variable
in the parameter estimation based on the selection of the best models.
Based on the MCMC chains, we subsequently determine the best values and
uncertainties for the stellar radius ($R$), the stellar age ($t$), and the mass of
the convective core ($M_\mathrm{cc}$) by extracting those quantities
from the model grid.

Before discussing the resulting distributions, we take a look at how well the grid modelling recovers the input parameters. Similarly to the above mentioned determined properties such as the stellar radius, we determined the values and uncertainties of $L/L_\sun$, $T_\mathrm{eff}$, and $\Pi_0$ from the best fitting model.

\begin{figure*}
    \centering
    \includegraphics[width=\textwidth]{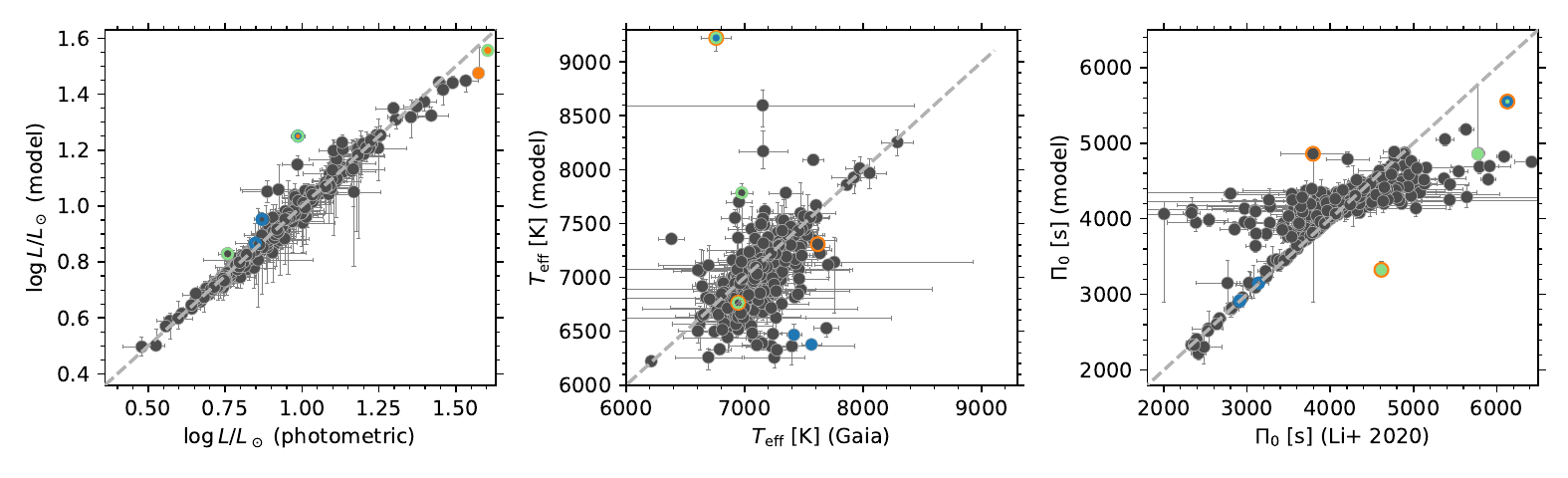}
    \caption{Correlations between observed properties (input values) and the recovered values (model). From left to right, we show $\log\,L/L_\sun$, $T_\mathrm{eff}$, and $\Pi_0$. All three properties have outlying values to a perfect recovery. Notably, outliers in one panel are not necessarily outliers in other panels. To show this property, we highlight the three most outlying points (difference divided by measurement uncertainty) in each panel with colour and indicate the same stars with coloured outlines in the other two panels. Outliers in luminosity are shown in orange, outliers in effective temperature in blue, and outliers in $\Pi_0$ in green. The shift towards intermediate $\Pi_0$ is discussed in the text.}
    \label{fig:recovered_values}
\end{figure*}

Figure~\ref{fig:recovered_values} shows the correlations between the input and recovered values. We find the majority of the recovered values to be in agreement with the input. In particular, the stellar luminosity was well retrieved. Yet, all three parameters have a varying level of outlying data points that could typically not be mapped to the grid during the  modelling. However, we note that only one star is an outlier in all three parameters. In most cases a single input parameter fell outside the grid space while the others gave good constraints, indicating possibly an inaccurate measurement of the one, outlying parameter.

A notably exception to the random outliers is the systematic shift towards intermediate $\Pi_0$ in the retrieved values. Stars for which a lower $\Pi_0$ was recovered than measured (i.e. stars to the right of the line of unity) have typically measured a very high value that would indicate young stars. Their position in the HRD (Fig.~\ref{fig:HRD}), places them well above the zero-age main sequence (ZAMS). Hence, these stars are either still on the pre-main sequence or well evolved with an inaccurate $\Pi_0$ measurement. Our grid concerns only the main sequence evolution which results in the latter assumption in the modelling. Revisiting these stars would be of great interest to either uncover populations of young stars among the \emph{Kepler} $\gamma$\,Dor stars or to rectify their asymptotic period-spacing. Stars to the left of the line of unity seem to be not as evolved in the HRD as indicated by their $\Pi_0$, yet we find no global reasoning that could be applied to all stars to explain their data. However, we note that stars for which we could not recover the luminosity and effective temperature are typically found at the edge of the instability strip in the HRD.

Quantitatively, the luminosity shows the strongest correlation between input and recovered values with a Pearson correlation $r=0.99$. The effective temperature is not that well recovered with $r=0.53$. Given the often inaccurate effective temperatures from the \emph{Gaia} catalogue this is not surprising. Despite the obvious horizontal band in the $\Pi_0$ comparison, we find $r=0.80$, due to the well recovered truly evolved stars with a low $\Pi_0$.

\begin{figure*}
    \centering
    \includegraphics[width=\textwidth]{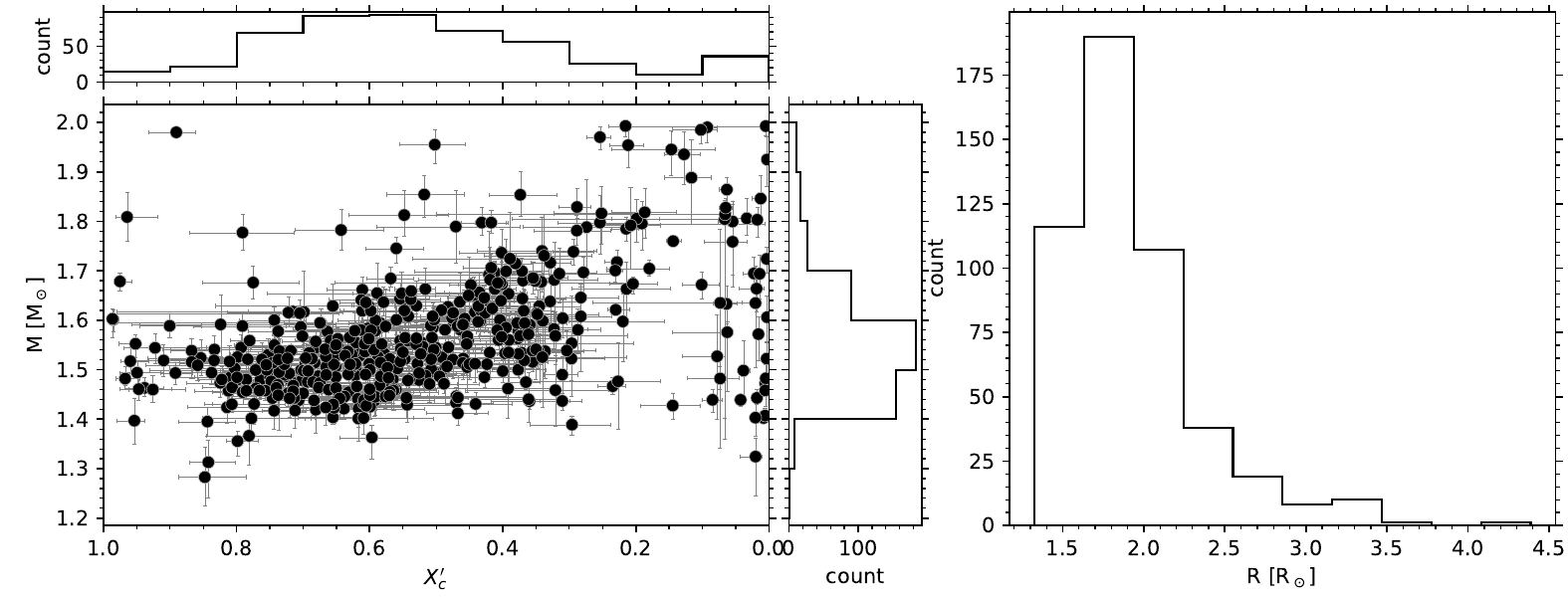}
    \caption{Distributions of the grid modelling results. Left: Asteroseismic mass against the normalised core hydrogen mass fraction $X_c'$. We find the majority of stars to have masses between 1.4 and 1.7\,$M_\odot$ and to be in the first part of their main sequence evolution with $X_c'>0.3$.
Note that the $X_c'$-axis is inverted to represent the time evolution from left to right. Right: Radius distribution of our sample. We find only a few evolved stars and most stars have a radius between 1.5 and 2.2 $R_\sun$ as expected in this mass regime for main sequence stars.}
    \label{fig:astero_results}
\end{figure*}

The asteroseismic masses, normalised core hydrogen mass fraction ($X_c'\equiv X_c/X_\mathrm{ini}$), and radii are shown in Fig.~\ref{fig:astero_results}. We find the majority of the stars to have $1.4\lesssim M/M_\sun \lesssim 1.7$ \citep[see also][]{Mombarg2024} and are a significant fraction into their main sequence evolution but not near their end ($X_c'>0.3$). The mean age of our sample is 1.3\,Gyr and we find indeed very few young stars (see the discussion above). The more massive stars in our sample tend to be most evolved. Overall, these distributions are in agreement with the position of the $\gamma$\,Dor instability strip (cf.\,Fig.~\ref{fig:HRD}) that is centred on these masses and narrows for evolved stars. Corresponding to the age and mass distributions, the stellar radii are distributed as expected for main sequence $\gamma$\,Dor stars with the majority in the range $1.5\lesssim R/R_\sun \lesssim 2.5$. However, the few evolved stars near the end of their main sequence life time have larger radii.

\begin{figure}
    \centering
    \includegraphics[width=\columnwidth]{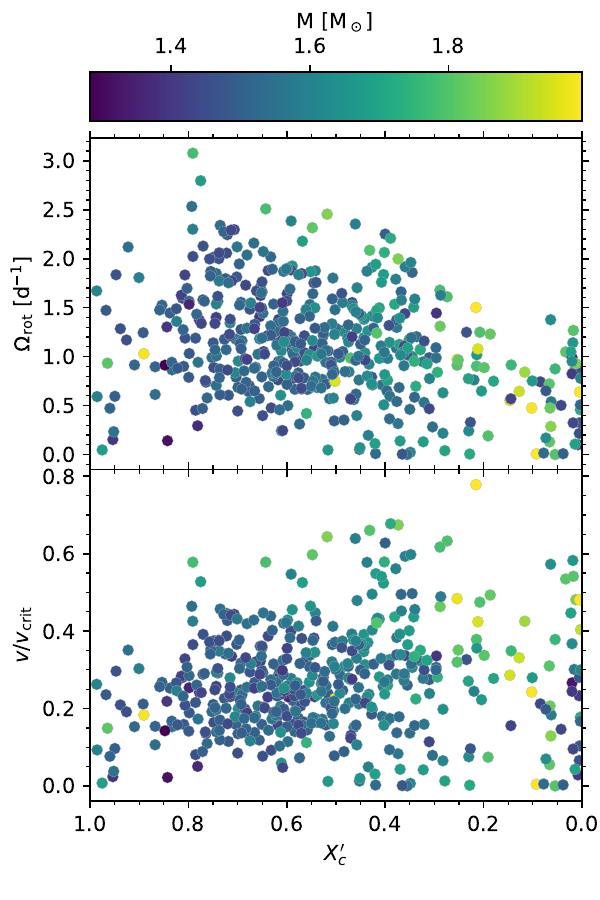}
    \caption{Rotational properties of the sample stars in the context of their evolutionary stage with the age represented by the normalised core hydrogen mass fraction ($X_c'$). We note the inverted x-axis. Top: The near-core rotation rate decreases with age as angular momentum is redistributed in the growing star. Bottom: The fraction of the critical rotation rate is mostly constant for the majority of stars with $0.1<$ \vvcrit$<0.6$.}
    \label{fig:rot_evol}
\end{figure}

Our $\gamma\,$Dor sample composed
in the previous section is the most homogeneous data set of red giant
progenitors with high-precision
asteroseismic observables on the main sequence, covering all possible rotation rates
\citep[cf.\,Fig.\,6 in][]{Aerts2021-RMP}.
As we are interested in the influence of rotation on stellar evolution, we now explore their second asteroseismic parameter, the near-core rotation rate $\Omega_\mathrm{rot}$. From Fig.~\ref{fig:rot_evol}, we find a decreasing near-core rotation rate with age as previously seen in \cite{GangLi2020} (as a function of $\Pi_0$).
Asteroseismology applied to stars with a convective core across
stellar evolution has shown that quasi-rigid rotation in the radiative envelope is an excellent approximation for the considered mass regime \citep{Aerts2019-ARAA,Aerts2021-RMP},
particularly for $\gamma\,$Dor stars \citep{VanReeth2018,GangLi2020,Saio2021} and can be modelled with an efficient angular momentum transport throughout the star \citep{Mombarg2023b,Moyano2024}.

Relying on the asteroseismic mass and radius, we estimate the current
ratio between the rotational velocity and the critical velocity
(\vvcrit) adopting the Keplerian approximation\footnote{For this estimation, we use the mass and radius from the non-rotating asteroseismic models.} as used in the MESA Isochrones \& Stellar Tracks \citep[MIST,][]{Dotter2016,Choi2016} models, which rely on the MESA version published by \citet{Paxton2015}. We find that very few stars in our sample exceed \vvcrit$=0.6$ (Fig.~\ref{fig:rot_evol}). Indeed the 95th percentile of the data is at \vvcrit$=0.53$, while 68\,\% of the stars are slower rotators with $0.15<$ \vvcrit$<0.39$. The upper boundary of \vvcrit increases with age but we observe only one $\gamma$\,Dor star with a near-critical rotation. In contrast to the trend of the near-core rotation, we find only a slight upward trend for the critical rotation rate with age while the majority of stars can be found at a constant rate \vvcrit$\approx 0.25$. This
result for field stars is in agreement with the recent findings for the young open cluster NGC\,2516 \citep{GangLi2024} in which the fastest $\gamma$\,Dor stars rotate at 0.4\,\vvcrit.

The results from our grid modelling, as well as the input parameters can be found in the online Table~\ref{tab:results}. Further, we provide \vvcrit and the derived MIST ages at certain evolutionary points.

\section{Age-dating the population of successors using MIST models}
\label{MIST_ages}

The main aim of this work is to provide age uncertainties at late
stages of stellar evolution by comparing non-rotating and rotating
stellar evolution tracks while keeping the other aspects of the
microscopic and macroscopic input physics the same. In order to make
this comparison, we rely on the often used MIST stellar evolution
tracks and isochrones \citep{Dotter2016,Choi2016}. Using precomputed
grids of isochrones to age-date evolved stars is a standard approach
in stellar evolution and Galactic archaeology studies.

The original database of MIST models in \citet{Choi2016} has been
extended to include rotation velocities ranging from zero velocity (no
rotation) to 90\,\% of the Keplerian critical velocity ($v_{\rm crit}$) for
increasing steps of 10\,\% in $v/v_{\rm crit}$ by
\citet{Gossage2019}. The entire MIST database is based on the same
input physics for all aspects of the models. It thus allows us to
assess the impact of rotation and its induced physical effects from
grid-based modelling.  The theory of shellular rotation and its
accompanying rotational mixing and gravity darkening adopted in the
MIST models is described extensively in \citet{Paxton2015}, to which
we refer for details.  We use the rotating MIST models from
\citet{Gossage2019} to estimate the effect of rotation on the ages of
stars in the mass range $M\in [1.3,2.0]\,$M$_\odot$ for their evolved
evolutionary phases lying ahead by using the population of
asteroseismically studied $\gamma\,$Dor main-sequence stars.

\begin{figure}
    \centering
    \includegraphics[width=\columnwidth]{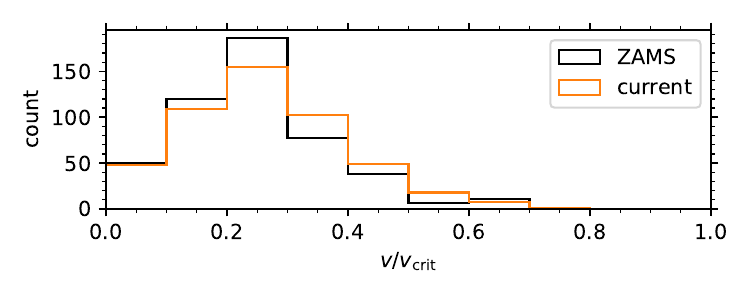}
    \caption{Distributions of rotation rates as fraction of the critical rotation (\vvcrit). The black distribution shows the back-traced initial rotation (at ZAMS). It is compared to the current distribution (orange) as derived from the asteroseismic grid modelling. The stars exhibit slightly larger fractions of the critical rotation velocity with respect to their initial rotation, as expected from the population seen in Fig.~\ref{fig:rot_evol}.}
    \label{fig:rot_distrib}
\end{figure}

Given the variation of the radius as a star evolves during the main
sequence, the critical rotational velocity and hence \vvcrit change
as the core hydrogen burning progresses. For each star we choose a family of tracks based on the maximum likelihood point estimator of the asteroseismic mass. Within this family, consisting of ten tracks ($0.0\leq$\vvcrit$\leq 0.9$ at ZAMS), we calculate \vvcrit at the grid point closest to the current core hydrogen mass fraction. Finally, we select the track closest to the observed $v/v_\mathrm{crit}$. Figure~\ref{fig:rot_distrib} shows the distribution of the back-traced (ZAMS) \vvcrit in comparison to the observed current distribution. The current distribution (as observed from the asteroseismic modelling) is broader and shifted slightly to faster rotation as it can also be seen from Fig.~\ref{fig:rot_evol}, when interpreted as an evolutionary sequence.

Once having selected the MIST track that comes closest to the
asteroseismic \vvcrit and age per star, its ages for future
evolutionary stages can easily be acquired by simply considering the
equivalent evolutionary point\footnote{The EEPs describe the stellar evolution in a time-independent way to simplify the comparison between models with different physical properties \citep[see][for more details]{Dotter2016}.} (EEP) on the MIST track.  We consider the age
effects at the terminal-age main sequence (TAMS, EEP 454), the tip of the red giant branch (TRGB,
EEP 605), the zero-age Helium-burning stage (ZAHeB, EEP 631), and the
terminal-age Helium-burning stage (TAHeB, EEP 707). For each of these
evolutionary points, we also extract the corresponding age from the
non-rotating track of the same mass as the reference value to estimate
the effect of internal rotation, by expressing it as a fraction of the
stellar age for the non-rotating case. Since rotation introduces extra
mixing in the stellar interior, we expect the ages of the stars in
evolved phases resulting from the rotating MIST tracks to be above
those of the non-rotating tracks.

\section{Age shifts in successor populations due to
  main-sequence rotation}
\label{sec:ageshifts}

The cores of most single stars born with $M\geq 1.2\,$M$_\odot$
generally rotate fast during core hydrogen burning because these stars
are not slowed down by magnetic braking. As proven by asteroseismology
of an ensemble of intermediate-mass field stars, their cores only slow
down considerably between the TAMS and the red giant phase
\citep[Fig.\,6 in][]{Aerts2021-RMP}. While it may be justified to
ignore the effects of rotation and internal mixing in stellar evolution models of
intermediate-mass stars after the TAMS, this is not a good
approximation during the initial 90\,\% of their evolution
\citep{Johnston2021}.
Despite this fact, age-dating of
red giants from asteroseismology is usually done by ignoring the
cumulative effect of the internal mixing caused by rotation in the
progenitor phases \citep[e.g.,][]{Martig2015,Ness2016,Basta2022}.
Moreover, the core overshooting description used to
mimic shear mixing in the core boundary layer is often
frozen to just one value in the models used to perform the
age-dating of red giants, instead of allowing it to vary during the
long main-sequence phase to mimic time-dependent
rotational shear mixing as done in this work.

While the latest age-dating methods, such as the one
developed by \citet{Basta2022}, can accommodate rotating models, its application to populations of red giants usually ignores the progenitor's rotational mixing,
underestimating in this way the age uncertainties.
Recent precision estimations for the ages of red giants with the best asteroseismology available quote $\sim\!\,$11\,\% age uncertainty when ignoring the rotation as of the ZAMS \citep{Montalban2021}.
Large population studies relying on a variety of CoRoT, {\it
  Kepler\/}, K2, and TESS data quote larger uncertainties up to some 25\,\%
\citep{Miglio2021,Stokholm2023,Willett2023}.

\begin{figure*}
    \centering\includegraphics[width=0.49\textwidth]{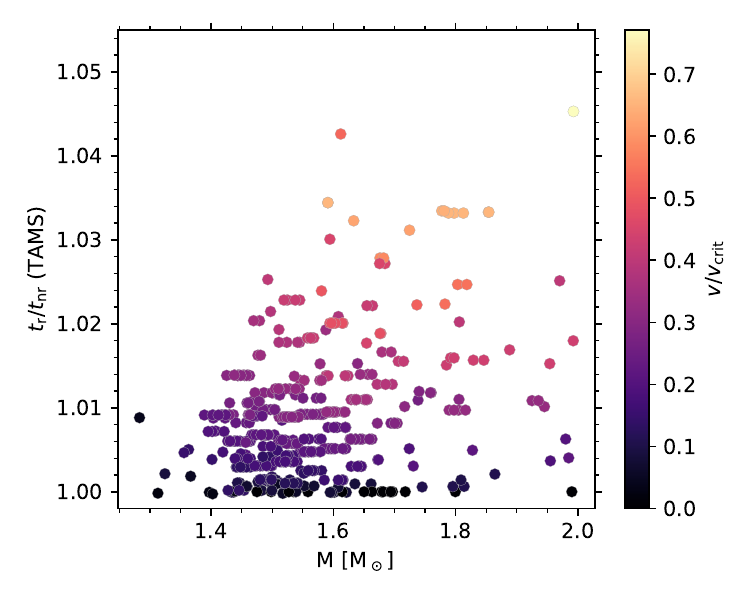}
    \centering\includegraphics[width=0.49\textwidth]{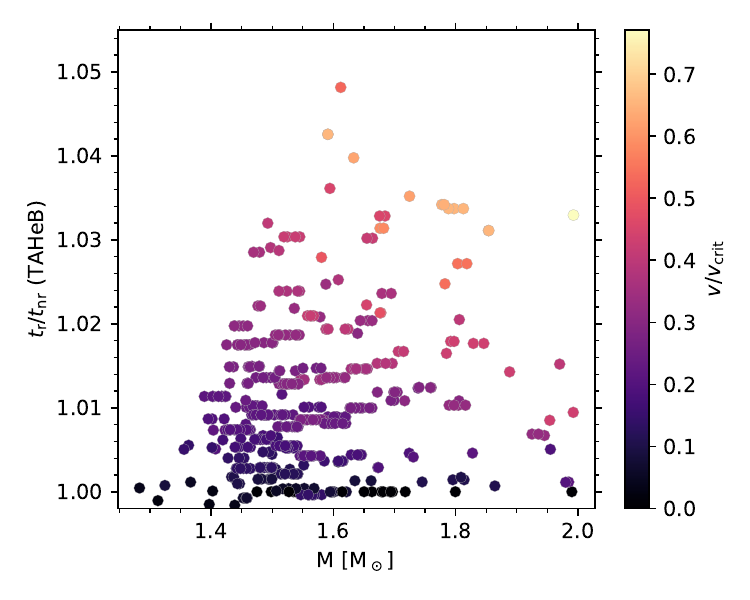}
    \caption{Relative ages between the rotating and non-rotating MIST tracks ($t_\mathrm{r}/t_\mathrm{nr}$) against stellar mass for two evolutionary points (TAMS left and TAHeB right). The colour-coding gives the ZAMS \vvcrit value. Both distributions are qualitatively similar although at the TAHeB the median value evolves to larger relative ages.}
    \label{fig:rel_ages}
\end{figure*}

Figure\,\ref{fig:rel_ages} shows the expected relative age differences at
the TAMS and the end of the core Helium burning stage (TAHeB) from the rotating and non-rotating MIST models
when we propagate the population of $\gamma\,$Dor stars forward
towards these evolutionary phases. It can be seen that the age
differences between rotating and non-rotating models reach up to 4.5\,\% at TAMS
for the fastest rotators but remain modest (below 1\,\%) for the
majority of stars for the physics of rotational mixing adopted in
the MIST models.

More specifically, we find a distinct upper envelope of the age spread
with stellar mass among the successors of the lowest-mass
$\gamma\,$Dor stars. The least massive evolved stars resulting from
the main-sequence sample ($M\simeq 1.3\,M_\sun$) are hardly affected
by the additional rotational mixing, given that their measured
rotation rates are close to those of a non-rotating star. Starting from
$M=1.4\,M_\sun$, the progenitor stars rotate typically between 5\,\%
and 20\,\% of the critical rotation, which adds $\le1$\,\% to the
stellar age by the end of the main sequence. For more
massive progenitor pulsators ($M\ge 1.5\,M_\sun$) the age difference
is purely a function of the rotation rate, that is, the distribution
of age differences follows the observed distribution of the
asteroseismic rotation rates and reaches $\sim$\!\,5\,\%. We point out that, as of
$M>1.7\,M_\sun$, the sample size of the $\gamma$\,Dor population is
small and may not be representative for all stars of such mass.

The distribution stays very similar at later evolutionary stages, although by the end of core Helium burning the overall age-spread between slowly and faster rotating $\gamma$\,Dor stars increases as seen from the right panel of Fig.~\ref{fig:rel_ages}. Notably the fastest rotating and most massive star exhibits a smaller relative age at TAHeB compared to the TAMS, contrary to other stars in the sample. The results for the two intermediate stages are shown in Appendix~\ref{app:ages}.

The histograms in Fig.~\ref{fig:age_hist} shows the evolution of the relative age distributions at the four considered evolutionary stages. As discussed above, the TAMS distribution shows the smallest relative age differences with a strong peak in the smallest bin, while the distributions are somewhat wider at later stages.

\begin{figure}
    \centering\includegraphics[width=\columnwidth]{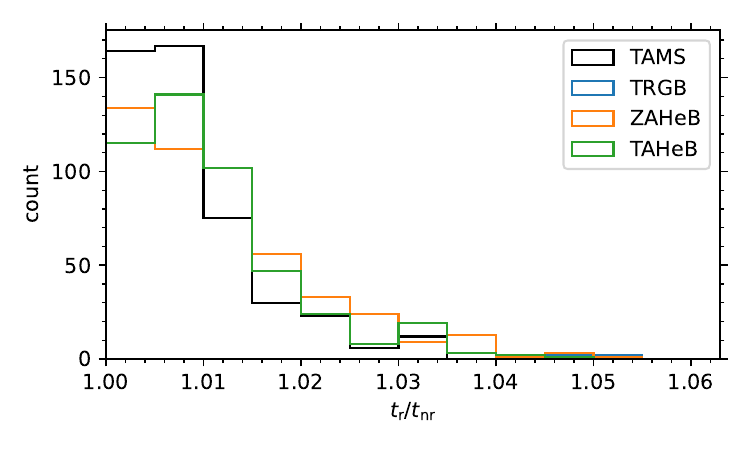}
    \caption{Histogram of the relative ages between rotating and non-rotating models at different evolutionary stages. The majority of red-giant ages are underestimated by a few percent under the assumption of non-rotating models. The TRGB distribution is indistinguishable from the distribution at ZAHeB and hence not visible.}
    \label{fig:age_hist}
\end{figure}

Our results reflect that rotational mixing is an important transport process throughout the long main sequence evolution, which lasts about 90\,\% of the entire lifetime of the star before it turns into a red giant and evolves further towards the white dwarf stage. Nevertheless, additional transport processes occurring specifically during the red giant phase (such as thermohaline mixing) and not considered here should also be
taken into account in the determination of ages for the
highest-precision Galactic archaeology based on surface abundances and
asteroseismology together \citep{Lagarde2017}.

\section{Additional systematic uncertainties}
\label{sec:systematics}
\subsection{For the asteroseismic parameters}

Recent computational work by \citet{Mombarg2024} has shown
that it is in principle possible to perform asteroseismic modelling of
intermediate-mass stars from rotating stellar models.
The 1D models delivered by \citet{Mombarg2022}
circumvent the spiky internal mixing profiles that occur when using the 1D prescriptions from \cite{Heger2000}, causing inaccuracies in the computation of oscillation modes.
\cite{Mombarg2022} rely on a 1D formalism of \cite{Zahn1992} for the implementation of rotational mixing in their models, while this is done self-consistently in 2D by \cite{Mombarg2023, Mombarg2024b}. In the 1D case, the rotational profile used for computing the efficiency of rotational mixing comes from models computed with the 2D ESTER evolution code \citep{EspinozaLara2013, Rieutord2016}. The 2D-to-1D
ESTER-informed mixing profile adopted in MESA evolution models brings
a potential improvement of major importance in asteroseismic modelling
of intermediate-mass stars with respect to the simplistic 1D treatment
of chemical mixing by means of a constant and ad hoc core overshooting description
as done in the literature \citep{Aerts2021-RMP} and adopted in
Sect.\,\ref{sec:model}.

Here, we provide grid-based modelling for our $\gamma\,$Dor sample
by using a 2D-to-1D model grid from \cite{Mombarg2024}, which differs from the previously employed MESA grid by the inclusion of angular momentum transport mechanisms. This rotational mixing replaces the constant envelope mixing ($D_\mathrm{mix}$) in the above used models. The angular momentum transport is implemented as diffusive processes and includes dynamical and secular shear instability, Eddington-Sweet circulation, Solberg-Høiland instability, Goldreich-Schubert-Fricke instability, and a Spruit-Tayler dynamo for angular momentum transport via magnetic torques. We refer the reader to \cite{Heger2000} and the MESA instrumentation papers \citep{Paxton2011, Paxton2013, Paxton2015, Paxton2018, Paxton2019} for details on these processes and their implementation in MESA. \cite{Mombarg2023b} calibrated the angular momentum transport from several slowly rotating, well-studied $\gamma$\,Dor stars with known near-core and surface rotation rates. Their new model grid is regularly sampled with $M \in [1.3, 2.0]$ ($\Delta M=0.1\,M_\sun$), $f_\mathrm{ov} \in [0.005, 0.035]$ ($\Delta f_\mathrm{ov}=0.005$), $\Omega/\Omega_\mathrm{crit} \in [0.05, 0.6]$ ($\Delta\Omega/\Omega_\mathrm{crit}=0.05$), and fixed a metallicity $Z=0.014$. The grid is more sparsely sampled than the non-rotating grid used above.

We applied these rotating models to our sample in the same way as outlined in Sects.\,\ref{sec:model}-\ref{sec:ageshifts} but included the measured asteroseismic near-core rotation rate as an additional input constraint. The stellar parameters obtained from this model are comparable to those from our previous analysis. On a star-to-star basis, shifts in the core hydrogen mass fraction are often compensated by changes in the core-overshoot and stellar mass. In combinations this results only in slight changes the evolution of the stellar properties with age as discussed in Fig.~\ref{fig:rot_evol}. Further, the initial distribution of rotation rates is similar to our previous results \citep[for a more detailed analysis of the initial rotation rates we refer the reader to][]{Mombarg2024}.

With only minor shifts in the overall asteroseismic stellar parameters compared to our initial model, in which the mixing due to rotation was modelled with additional core overshoot, we do not find significant difference for the MIST-based ages at later evolutionary stages with respect to the analysis above. Some stars have slightly different relative ages at the four considered evolutionary points, yet the overall distribution hardly changes. This illustrates that mimicking element transport processes by a fudge factor in the overshoot zone is a valid approach, as long as the chemical mixing can be described diffusively \citep[cf.,][]{Aerts2021-RMP}.

\subsection{From evolution models with meridional circulation}

In contrast to the MIST models computed with the code MESA, which adopts a diffusive approach and treats the
rotationally induced mixing and angular momentum processes as described in \citet{Heger2000},
other stellar evolution models rely on a diffusive-advective treatment. Such a treatment allows for the inclusion of meridional circulation, often implemented following the theory by \citet{Maeder1998}. Here we rely on the publicly available Geneva models computed by
\citet{Georgy2013}, in order to assess the systematic age uncertainty caused by the different input physics to treat rotation.

\citet{Ouazzani2019} have shown from asteroseismology of $\gamma\,$Dor stars that their diffusive-advective models computed with the CESTAM code, which also relies on the theory by \citet{Maeder1998}, need fixes in terms of angular momentum transport.  Better agreement between models and asteroseismic data has been achieved by
\citet{Moyano2023,Moyano2024} from the inclusion of internal magnetism in the Geneva code, in addition to meridional circulation and shear mixing. However, neither these new Geneva models nor the CESTAM ones are publicly available. For this reason, we assess the age uncertainty due to absence or inclusion of meridional circulation by comparing the results from the MIST models with those derived from the SCYLIST Geneva models made publicly available by \citet{Georgy2013}. The use of these two sets of isochrones and stellar tracks is common to age-date stars.

 It has already been shown that the inclusion of meridional circulation can result in significant differences of the evolutionary tracks of rotating models beyond the TAMS \citep{Martins2013}. The influence of rotation on luminosity and temperature are analysed in detail in \cite{Choi2016} and \cite{Gossage2018}. We deduce from Fig.\,20 of \cite{Choi2016} that the ages between rotating MIST models and Geneva models are quite different for stars with $M<3\,\mathrm{M}_\sun$, the mass range considered in this work.

For the models with a mass of 1.7\,M$_\sun$, which is the lowest mass in the publicly available grid of rotating Geneva models computed by \citet{Georgy2013}, we find a relative age difference of 10\,\% for \vvcrit$=0.2$
at the TRGB compared to the non-rotating model, while this difference increases to 20\,\% for \vvcrit$=0.4$. These values are significantly larger than in the MIST case, for which we found 1\,\% and 2\,\%, respectively. When comparing ages obtained from the MIST models to these from the Geneva models, we find a very similar TRGB age in the non-rotating case, albeit a smaller TAMS age. Yet, in the Geneva case the main sequence life time is extended by $\sim\!30\,\%$ for \vvcrit$=0.4$, resulting in a large relative age difference at the TRGB.

We conclude that different input physics for angular momentum and/or element transport, notably the inclusion of meridional circulation due to rotation or not,
may add significant systematic uncertainties in red giant ages when taking into account the rotation in their progenitor phases.

\subsection{Sample selection}

For this study, we employed the largest homogeneous sample of $\gamma$\,Dor stars with $\Pi_0$ and $\Omega_{\rm rot}$ estimates
available in the literature, in order to assess the influence of rotation on the age distribution of red giants. Despite being the largest such sample, it is still small compared to the overall stellar population and selection effects might influence our results.

The \emph{Kepler} field contains mostly older field stars \citep{Berger2020}, as also seen from Fig~\ref{fig:astero_results}. Hence, we are missing the youngest $\gamma$\,Dor stars in our analysis and our sample might not represent the full variety of $\gamma$\,Dor stars. Future, large-scale studies similar to \cite{GangLi2020} with stars observed with the Transiting Exoplanet Survey Satellite (TESS) might provide insights into a younger population of $\gamma$\,Dor stars since the TESS continuous viewing zone contains a much younger population of stars than the \emph{Kepler} field \citep{Avallone2022}.

\cite{Aerts2023} present distributions of stellar parameters for a much larger sample of 11\,636 $\gamma$\,Dor star candidates identified from \emph{Gaia} photometry \citep{DeRidder2023}. Our distributions of stellar parameters ($T_\mathrm{eff}$, $\log L/L_\sun$, $R/R_\sun$) are in qualitative agreement with these results. The majority of these stars are currently being analysed in detail from high-cadence TESS space photometry to confirm their $\gamma\,$Dor nature (Hey \& Aerts, in prep.) but we do not yet possess asteroseismic parameters such as $\Pi_0$ and $\Omega_\mathrm{rot}$ for them. The stars from this sample (next to stars in the TESS continuous viewing zones) are prime targets for future population studies with TESS.

\subsection{Insights from more massive stars}

Since we focused our analysis on $\gamma$\,Dor stars, our sample has a very limited mass-range. The results of our grid modelling place most of the stars in the range 1.4\,M$_\odot\leq$\,M\,$\leq 1.7\,$M$_\odot$, which coincides with the centre of the $\gamma$\,Dor instability strip. As seen from Fig.~\ref{fig:rot_evol} the stars in our sample are mostly moderate rotators while higher-mass stars often rotate at higher fractions of the critical velocity. Similar to the sample of $\gamma$\,Dor stars, \cite{Pedersen2021} homogeneously analysed a sample of 26 slowly pulsating B (SPB) stars from the \textit{Kepler} mission with asteroseismic forward modelling.

We placed the updated results for this sample from \citet{Pedersen2022a} on the MIST tracks in a similar fashion to our analysis above.
Since we have only a small sample, we cannot deduce a statistically meaningful result. Yet, we find a similar relative age difference of $2-3\,\%$ for the majority of slowly rotating stars (\vvcrit$\leq0.5$). However, near-critical rotation is a common phenomenon among B stars \citep{Townsend2004} and some stars in the sample of \cite{Pedersen2022a} rotate with \vvcrit$>0.7$. For these stars the main sequence lifetime is extended by up to 50\,\% at $M\approx 4\,M_\sun$. While this is a higher mass than for most red giant stars, we have seen that the prolonged main sequence for the more massive $\gamma$\,Dor stars is mainly a function of the rotation rate and not so much of the mass.
Hence, the distribution of the rotation rates of lower-mass SPB stars might be directly transferable to a distribution of age uncertainties for the mass regime $[2,4]\,$M$_\odot$.

\section{Conclusions}

Based on \nsample{} $\gamma$\,Dor stars with homogeneously deduced astrophysical parameters from Gaia DR3 and 4-year {\it Kepler\/} light curves, we characterise the population of red giant progenitors with masses in the range $1.3\le M/M_\sun\le 2.0$. We employ asteroseismic grid modelling and obtain masses, the current core hydrogen mass fraction, and radii for these stars. Together with the asteroseismic near-core rotation rate, we adopt a common approach to age-date stars and place the sample stars on rotating MIST evolutionary tracks.

The current population of $\gamma$\,Dor stars show moderate rotation with $0.15\lesssim$ \vvcrit$\lesssim0.4$ for the majority of stars. When comparing the influence of this rotation to age-dating at later stages of their evolution, we find a modest increase of the age. However, the fastest rotators can be up to 5\,\% older when leaving the main sequence. Typical fast rotation (\vvcrit$\sim\!0.4$) increases the main sequence age by 2\,\% when relying on the transport processes adopted in the MESA stellar evolution code, but can reach up to 20\% when relying on models including meridional circulation.

Given the unknown rotation history of isolated red giants in the Milky Way, the age spreads caused by their rotating progenitors as presented here should always be considered in the uncertainty analysis for (asteroseismic) red giant ages. Effects of core boundary mixing during core helium burning are nowadays taken into account in the construction of red giant models calibrated by asteroseismology \citep{Constantino2015,Bossini2017,Noll2024}. Yet the rotational mixing during the long main-sequence phase is usually ignored, while it has a major impact on the size and mass of the helium core by the TAMS \citep{Johnston2021,Pedersen2022a}. As we have shown here the rotation during the main sequence also impacts the age by the time the star is a red giant.
Our work shows the importance of taking into account transport processes in the stellar interior during the entire main sequence when age-dating red giants for high precision Galactic archaeological studies.
Asteroseismology of young and medium-aged open clusters with a variety of metallicities and both main sequence and red giant pulsators among their members offers the best way to calibrate internal mixing and angular momentum transport processes across the entire nuclear evolution phases of intermediate-mass stars
\citep[see][for initial attempts of such modelling]{Fritzewski2024,GangLi2024}.

\begin{acknowledgements}
The authors appreciate the swift report and encouraging words from the anonymous referee.
    CA is grateful to Melissa Ness and Danny Horta for pleasant and valuable discussions on Galactic archaeology. The research leading to these results has received funding from    the KU\,Leuven Research Council (grant C16/18/005: PARADISE).  CA
    also acknowledges financial support from the European Research
    Council (ERC) under the Horizon Europe programme (Synergy Grant
    agreement N$^\circ$101071505: 4D-STAR). While partially funded by the European Union, views and opinions expressed are however those of the author(s) only and do not necessarily reflect those of the European Union or the European Research Council. Neither the European Union nor the granting authority can be held responsible for them.
    JSGM acknowledges funding from the French Agence Nationale de la Recherche (ANR), under grant MASSIF (ANR-21-CE31-0018-02).
    This research has made use of NASA's Astrophysics Data System
    Bibliographic Services and of the SIMBAD database and the VizieR
    catalogue access tool, operated at CDS, Strasbourg, France.
    This work has made use of data from the European Space Agency
    (ESA) mission \emph{Gaia} (\url{https://www.cosmos.esa.int/gaia}),
    processed by the \emph{Gaia} Data Processing and Analysis
    Consortium (DPAC,
    \url{https://www.cosmos.esa.int/web/gaia/dpac/consortium}). Funding
    for the DPAC has been provided by national institutions, in
    particular the institutions participating in the \emph{Gaia}
    Multilateral Agreement.
    \newline
    \textbf{Software:}
    This work made use of \texttt{Topcat} \citep{2005ASPC..347...29T}.
    This research made use of the following \texttt{Python} packages:
    \texttt{IPython} \citep{ipython};
    \texttt{MatPlotLib} \citep{Hunter:2007};
    \texttt{NumPy} \citep{numpy2020};
    \texttt{Pandas} \citep{pandas};
    \texttt{SciPy} \citep{scipy}.

\end{acknowledgements}

\bibliographystyle{aa} 
\bibliography{DarioConny.bib} 


\begin{appendix}
\section{Online table}

Table~\ref{tab:results} gives an overview on the columns of the online Table.

\begin{table*}[]
    \caption{Overview on the columns in the online table containing the input parameters and asteroseismically determined parameters for the \nsample{} analysed $\gamma$\,Dor stars.}
    \begin{tabular}{l l l}
    \hline
    \hline
        Name & Unit & Description \\
        \hline
        KIC & - & KIC ID\\
        RA & deg & Right ascension from \emph{Gaia} DR3\\
        Dec & deg & Declination from \emph{Gaia} DR3\\
        Lum & $L_\sun$ & Photometric Luminosity ($L$)\\
        err\_Lum & $L_\sun$ & Uncertainty on photometric Luminosity\\
        Teff& K & Effective temperature from \emph{Gaia} DR3 ($T_\mathrm{eff}$)\\
        err\_Teff& K & Uncertainty on effective temperature\\
        Pi0& s & Buoyancy travel time from \cite{GangLi2020} ($\Pi_0$)\\
        errl\_Pi0 & s & Lower uncertainty on $\Pi_0$ from \cite{GangLi2020}\\
        erru\_Pi0 & s & Upper uncertainty on $\Pi_0$ from \cite{GangLi2020}\\
        frot & d$^{-1}$ & Near-core rotation rate from \cite{GangLi2020} ($\Omega_\mathrm{rot}$)\\
        errl\_frot & d$^{-1}$ & Lower uncertainty on $\Omega_\mathrm{rot}$ from \cite{GangLi2020}\\
        erru\_frot & d$^{-1}$ & Upper uncertainty on $\Omega_\mathrm{rot}$ from \cite{GangLi2020}\\
        Xc' & - & Normalised core hydrogen mass fraction ($X_c'$)\\
        errl\_Xc' & - & Lower uncertainty on $X_c'$\\
        erru\_Xc' & - & Upper uncertainty on $X_c'$\\
        age & Gyr & Asteroseismic age ($t$)\\
        errl\_age & Gyr & Lower uncertainty on age\\
        erru\_age & Gyr & Upper uncertainty on age\\
        mass & $M_\sun$ & Asteroseismic mass ($M$)\\
        errl\_mass & $M_\sun$ & Lower uncertainty on mass\\
        erru\_mass & $M_\sun$ & Upper uncertainty on mass\\
        radius & $R_\sun$ & Asteroseismic radius ($R$)\\
        errl\_radius & $R_\sun$ & Lower uncertainty on radius\\
        erru\_radius & $R_\sun$ & Upper uncertainty on radius\\
        vvcrit & - & Ratio of current rotation rate to critical rotation rate (\vvcrit)\\
        MIST\_age\_TAMS & Gyr & Age of rotating MIST track at TAMS ($t_\mathrm{r}$)\\
        NR\_MIST\_age\_TAMS & Gyr & Age of non-rotating MIST track at TAMS ($t_\mathrm{nr}$)\\
        TAMS\_frac & - & Fraction of rotating to non-rotating age at TAMS ($t_\mathrm{r}/t_\mathrm{nr}$)\\
        MIST\_age\_TRGB & Gyr & Age of rotating MIST track at TRGB ($t_\mathrm{r}$)\\
        NR\_MIST\_age\_TRGB & Gyr & Age of non-rotating MIST track at TRGB ($t_\mathrm{nr}$)\\
        TRGB\_frac & - & Fraction of rotating to non-rotating age at TRGB ($t_\mathrm{r}/t_\mathrm{nr}$)\\
        MIST\_age\_ZAHeB & Gyr & Age of rotating MIST track at ZAHeB ($t_\mathrm{r}$)\\
        NR\_MIST\_age\_ZAHeB & Gyr & Age of non-rotating MIST track at ZAHeB $t_\mathrm{nr}$)\\
        ZAHeB\_frac & - & Fraction of rotating to non-rotating age at ZAHeB ($t_\mathrm{r}/t_\mathrm{nr}$)\\
        MIST\_age\_TAHeB & Gyr & Age of rotating MIST track at TAHeB ($t_\mathrm{r}$)\\
        NR\_MIST\_age\_TAHeB & Gyr & Age of non-rotating MIST track at TAHeB ($t_\mathrm{nr}$)\\
        TAHeB\_frac & - & Fraction of rotating to non-rotating age at TAHeB ($t_\mathrm{r}/t_\mathrm{nr}$)\\
        \hline
    \end{tabular}

    \label{tab:results}
\end{table*}

\FloatBarrier

\section{Comparison between asteroseismically informed grid modelling and the \emph{Gaia}-FLAME results}

\begin{figure*}
    \centering
    \includegraphics[width=\textwidth]{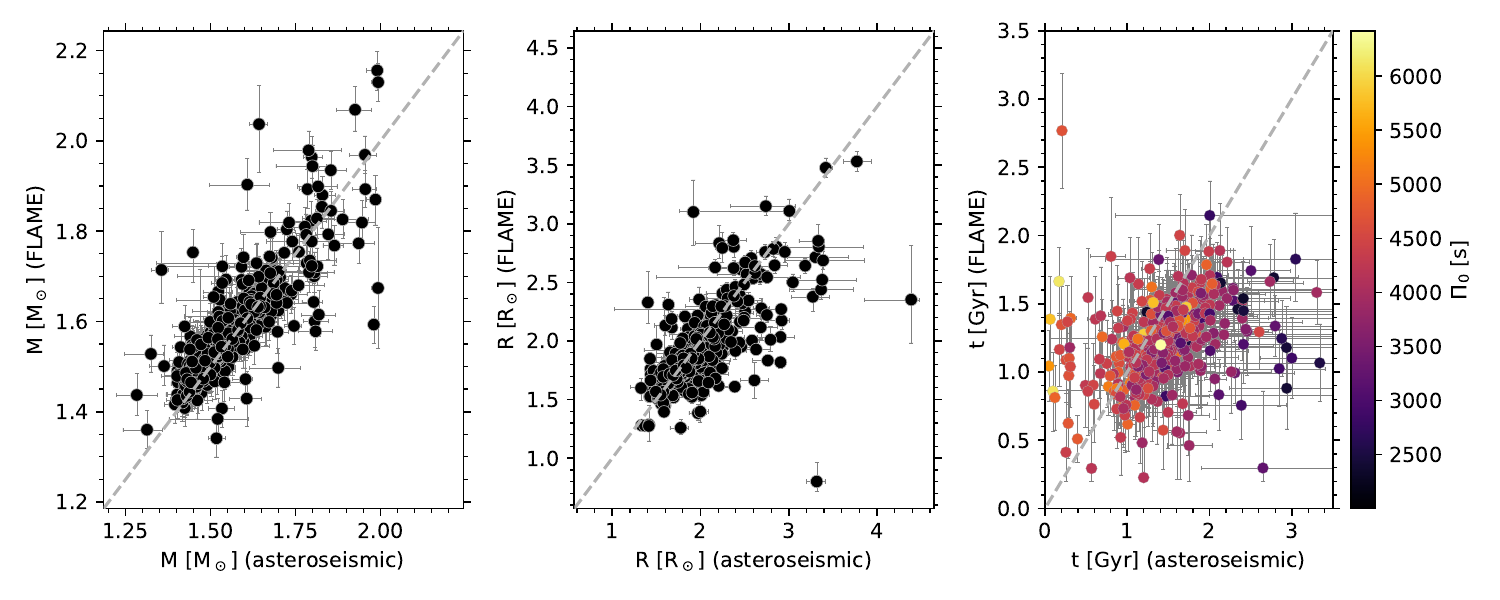}
    \caption{Comparison between asteroseismic parameter estimates and the \emph{Gaia}-FLAME values. Left: The stellar mass estimate from our analysis is mostly consistent with the FLAME pipeline. However, a slight systematic offset to lower masses in our analysis exist. Centre: Most radii agree between the two methods, while some outliers with large radii from the asteroseismic analysis can be found. These stars have typically a larger asteroseismic age and are more evolved in our analysis. Right: The additional asteroseismic information from $\Pi_0$ gives a good handle on the stellar age. Consequently, we observe a large scatter and larger spread in age compared to \emph{Gaia}-FLAME.}
    \label{fig:flame_comp}
\end{figure*}

The \emph{Gaia}-FLAME parameters in the astrophysical parameter tables of \emph{Gaia} DR3 do not only include the above mentioned luminosity but also estimates of the stellar mass, radius, and age for most of the observed stars. In Fig.~\ref{fig:flame_comp}, we show these three parameters in comparison to our asteroseismically inferred values. Both the masses and the radii are overall in good agreement, while the ages estimated from both methods diverge.

Despite being in good agreement, we find that on average the asteroseismic mass is slightly lower than the FLAME mass. This offset can mainly be attributed to the different underlying stellar models. The outliers in the left panel of Fig.~\ref{fig:flame_comp} are at the same level as seen in previous comparisons in the main text.

The radii are also in good agreement between the two methods. However, we find more outliers with larger asteroseismic radii. These are stars are likely more evolved in our model.

The ages are not in a good agreement between our asteroseismically informed model and FLAME.
While the ages from the FLAME pipeline are mostly constrained between 0.5 and 2\,Gyr, our results are between 0 and 3\,Gyr. The colour gradient in right panel of Fig.~\ref{fig:flame_comp} shows that the additional constraint from $\Pi_0$ has great probing power for stellar ages.

\section{Age spread at different evolutionary points}

\begin{figure*}
    \includegraphics[width=0.49\textwidth]{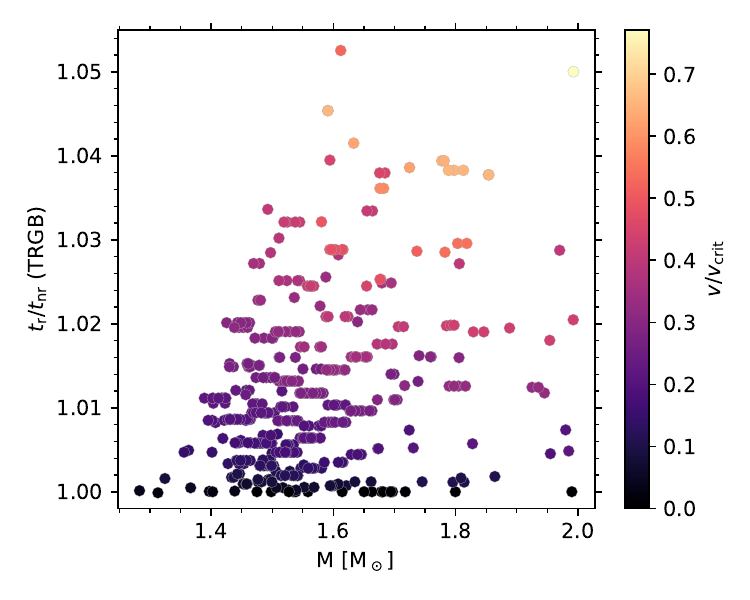}
    \includegraphics[width=0.49\textwidth]{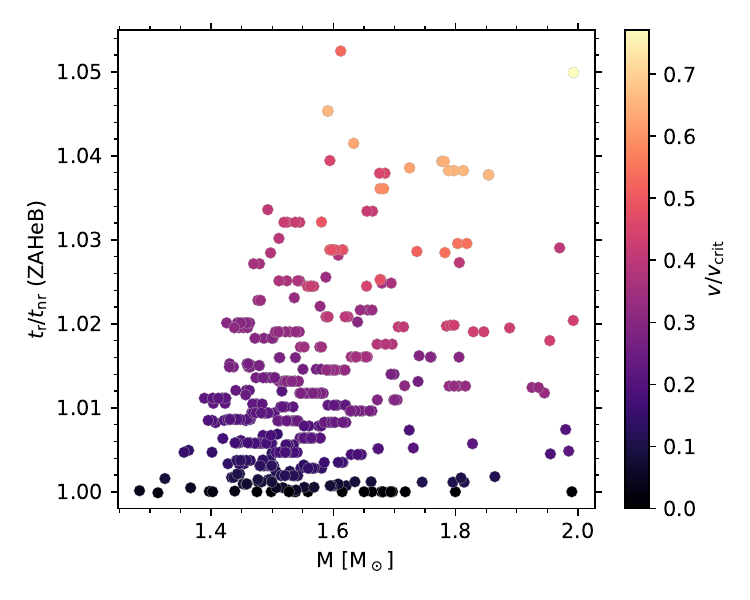}
    \caption{Relative ages at TRGB (left) and the ZAHeB (right) comparing rotating and non-rotating
      models similar to Fig~\ref{fig:rel_ages}.}
    \label{fig:trgb}
\end{figure*}

We analysed the relative ages not only at the end of the core hydrogen and core helium burning but also at the tip of the red giant branch and the beginning of the core helium burning. Fig.~\ref{fig:trgb} shows the distributions for these two evolutionary stages. As already seen from the histogram in Fig.~\ref{fig:age_hist} the distributions are very similar with a very slight evolution between them.

\label{app:ages}

\end{appendix}

\end{document}